\documentclass[preprint,preprintnumbers,amsmath,amssymb]{revtex4}
\usepackage{graphicx}
\usepackage{bm}

\begin{document}

\title{Dynamics of quantum discord in asymmetric and local non-Markovian environments}
\author{Xiang Hao}
\altaffiliation{Corresponding author,Email:110523007@suda.edu.cn}

\author{Tao Pan}

\author{Jinqiao Sha}

\affiliation{Department of Physics, School of Mathematics and
Physics, Suzhou University of Science and Technology, Suzhou,
Jiangsu 215011, People's Republic of China}

\author{Shiqun Zhu}

\affiliation{School of Physical Science and Technology, Suzhou
University, Suzhou, Jiangsu 215006, People's Republic of China}

\begin{abstract}

The non-Markovian decoherence of quantum and classical correlations
is analytically obtained when two qubits are asymmetrically
subjected to the bit flip channel and phase flip channel. For one
class of initial mixed states, quantum correlations quantified by
quantum discord decay synchronously with classical correlations. The
discovery that the decaying rates of quantum and classical
correlations suddenly change at the characteristic time is
physically interpreted by the distance from quantum state to the
closest classical states. In a large time interval, quantum
correlations are greater than classical correlations. The quantum
and classical correlations can be preserved over a longer period of
time via the kernel characterizing the environment memory effects.



\end{abstract}

\maketitle

\section{Introduction}

The great obstacle to the practical tasks of quantum information
processing is the destruction of quantum correlations arising from
the inevitable interactions of the systems of interest with their
environments \cite{Nielsen00,Zurek03}. The fundamental quantum
correlations are contained in nonclassical states. It is known that
the entanglement measure can only quantify quantum correlations for
inseparable states. Recently, a large amount of experimental and
theoretical studies have discovered that quantum discord
\cite{Ollivier01,Henderson01} is a more general measure of
nonclassical correlations in comparison with the entanglement
\cite{Horodecki09,Datta08,Lanyon08,Luo08,Sarandy09,Lutz09,Piani08,Modi10,Mazhar10,Werlang09,Fanchini10,Wang10,Werlang10,Maziero09,Maziero10,Celeri10,Mazzola10,Xu10,He10,Chen10,Guo10}.
Quantum correlation introduced by Henderson and Vedral
\cite{Henderson01} is identical to the quantum discord defined by
Ollivier and Zurek \cite{Ollivier01} in the case of two qubits
\cite{Guo10}. Specially, the separable mixed states with nonzero
quantum discord are also considered as one kind of useful resources
contributing to the speedup of some quantum computations. Therefore,
the dynamics of quantum correlations have received much attention.
Until now, the Markovian and non-Markovian decoherences of
correlations have been investigated in Refs.
\cite{Fanchini10,Wang10,Werlang10,Maziero09,Maziero10,Celeri10,Mazzola10,Xu10}.
It is found that the exponential decay of quantum discord is
distinct from the sudden death of the entanglement \cite{Yu04} at
finite time in the Markovian environments. The decline of quantum
discord can be delayed in the non-Markovian channels
\cite{Fanchini10}. But quantum correlations are usually less than
classical correlations in a large time interval. In many interesting
studies, it is assumed that qubits are coupled to local reservoirs
which are similar to each other. However, the properties of local
environments are often different from each other. This means that
the practical environments are asymmetric. The asymmetric
characterizations of the decoherence channels for every qubit must
bring about the positive or negative effects on the behavior of
quantum and classical correlations.

In this paper, we study the dynamics of quantum and classical
correlations in the influence of asymmetric and non-Markovian local
environments. For the spin system consisting of two noninteracting
qubits $A$ and $B$, the non-Markovian decoherence channel of each
qubit is given by the post-Markovian phenomenological approach in
Ref. \cite{Shabani05}. The effects of the bit-flip channel and
phase-flip channel are taken into account. In section II, the time
evolution of quantum states is analytically obtained when a general
class of Bell-diagonal mixed states are chosen to be initial ones.
In section III, three different types of the behavior of quantum
discord and classical correlations are deduced with the change of
initial states. A simple discussion concludes the paper.

\section{The evolution under asymmetric decoherence channels}

The Shabani and Lidar post-Markovian master equation
\cite{Shabani05} is used to describe the time evolution of quantum
bipartite states $\rho(t)$ in the non-Markovian environments. The
general form of the master equation is expressed by
\begin{equation}
\dot{\rho}(t)=\hat{L} \int_{0}^{t}dt'k(t')\exp(t'\hat{L})\rho(t-t'),
\end{equation}
where the operator $\hat{L}$ is the Markovian Liouvillian and $k(t)$
is the kernel characterizing the environment memory effects.
Specially, $k(t)=\delta(t)$ in the Markovian case. For a given
system and environment, the kernel can be experimentally engineered
via quantum state tomography \cite{Shabani05}. For the Pauli
channels, $\hat{L}\rho(t)=\sum_{j=A,B}\frac
a2(2\sigma^{j}_{\alpha}\rho\sigma^{j}_{\alpha}-\sigma^{j}_{\alpha}\sigma^{j}_{\alpha}\rho-\rho\sigma^{j}_{\alpha}\sigma^{j}_{\alpha})$
where $\alpha=x,y,z$ corresponds to the bit flip, the bit-phase and
phase flip channel respectively. The parameter $a$ represents the
decaying rate. According to Ref. \cite{Tan10}, the post-Markovian
channels are adopted to describe the actual dynamics approximately.
For simplicity, qubits $A$ and $B$ are subjected to the two
different channels of bit flip $\varepsilon_{BF}$ and phase flip one
$\varepsilon_{PF}$ respectively. The bit flip channel for the $j-th$
qubit can be given by $ \varepsilon_{BF}^{j}(|e\rangle_j\langle
e|)=\frac 12[I^{j}+p(t)\sigma^{j}_{z}]$, $
\varepsilon_{BF}^{j}(|e\rangle_j\langle g|)=\frac
12[\sigma^{j}_{x}+ip(t)\sigma^{j}_{y}]$, $
\varepsilon_{BF}^{j}(|g\rangle_j\langle e|)=\frac
12[\sigma^{j}_{x}-ip(t)\sigma^{j}_{y}]$ and
$\varepsilon_{BF}^{j}(|g\rangle_j\langle g|)=\frac
12[I^{j}-p(t)\sigma^{j}_{z}]$.  Here $|e(g)\rangle_{j}$ is the
eigenstate of $\sigma^{j}_{z}$ with the corresponding eigenvalue
$\pm 1$ and $I^{j}$ is the identity operator of qubit $j$. If the
kernel is assumed to be $k(t)=A\exp(-\gamma t)$, the parameter
$p(t)$ is decided by $ p(t)=\exp(-\frac
{2a+\gamma}{2}t)(\cos\omega_{0}t+\frac
{2a+\gamma}{2\omega_{0}}\sin\omega_{0}t)$ where
$\omega_{0}=\sqrt{2aA-[(2a+\gamma)/2]^2}$. The phase flip channel
can also be defined by $\varepsilon_{PF}^{j}(|e\rangle_j\langle
e|)=\frac 12(I^{j}+\sigma^{j}_{z})$,
$\varepsilon_{PF}^{j}(|e\rangle_j\langle g|)=\frac
{p}2(\sigma^{j}_{x}+i\sigma^{j}_{y})$,
$\varepsilon_{PF}^{j}(|g\rangle_j\langle e|)=\frac
{p}2(\sigma^{j}_{x}-i\sigma^{j}_{y})$ and
$\varepsilon_{PF}^{j}(|g\rangle_j\langle g|)=\frac
12(I^{j}-\sigma^{j}_{z}).$

The initial states of two qubits are one important Bell-diagonal
mixed states in the general form of
 \begin{equation}
\rho^{AB}(0)=\frac 14 \left(I^{A}\otimes
I^{B}+\sum_{\alpha=x,y,z}c_{\alpha}\sigma^{A}_{\alpha}
\sigma^{B}_{\alpha}\right),
\end{equation} where the real number
satisfies $0\leq |c_{\alpha}|\leq 1$. These symmetric states are
ones with maximally mixed marginals, i. e.,
$\mathrm{Tr}_{A(B)}\propto I^{B(A)}$. Under the two asymmetric
decoherence channels, quantum bipartite states at any time can be
analytically calculated by
\begin{equation}
\rho^{AB}(t)=\varepsilon_{BF}^{A}\otimes
\varepsilon_{PF}^{B}[\rho^{AB}(0)].
\end{equation} The expression of $\rho^{AB}(t)$ is given by
\begin{align}
\rho^{AB}(t)=&\lambda_{1}(t)|\Psi^{+}\rangle \langle
\Psi^{+}|+\lambda_{2}(t)|\Phi^{+}\rangle \langle \Phi^{+}|& \\
\nonumber &+\lambda_{3}(t)|\Phi^{-}\rangle \langle
\Phi^{-}|+\lambda_{4}(t)|\Psi^{-}\rangle \langle \Psi^{-}|,&
\end{align}
where the four Bell states are defined by $|\Psi^{\pm}\rangle=\frac
1{\sqrt{2}}(|eg \rangle \pm |ge \rangle)$ and
$|\Phi^{\pm}\rangle=\frac 1{\sqrt{2}}(|ee \rangle \pm |gg \rangle)$.
The coefficients $\lambda_{i}(t)$ are written by$
\lambda_{1,2}(t)=\frac 14[1+c_{x}(t)\pm c_{y}(t)\mp c_{z}(t)]$ and
$\lambda_{3,4}(t)=\frac 14[1-c_{x}(t)\mp c_{y}(t) \pm c_{z}(t)]$
where $c_{x}(t)=p(t)c_{x}$, $c_{y}(t)=p^2(t)c_{y}$ and
$c_{z}(t)=p(t)c_{z}$. It is shown that quantum states also keep the
symmetric Bell-diagonal expression under the asymmetric Pauli
channels. The mixing proportions of Bell states are varied in time.

\section{The behavior of quantum and classical correlations}

To clearly describe the dynamics of correlations, we use quantum
discord as one general measure for quantum correlation. For an
arbitrary bipartite state $\rho^{AB}$, the total correlations are
expressed by quantum mutual information $
I(\rho^{AB})=\sum_{j=A,B}S(\rho^{j})-S(\rho^{AB}),$ where $\rho^{j}$
represents the reduced density matrix of subsystem $j$ and
$S(\rho)=-\mathrm{Tr}(\rho \log_{2}\rho)$ is the von Neumann
entropy. Henderson and Vedral \cite{Henderson01} proposed one
measure of bipartite classical correlation $C(\rho^{AB})$ based on a
complete set of local projectors $\{ \Pi^{B}_{i}\}$ on one subsystem
$B$. After the local measurements, the reduced state of subsystem
$A$ can be written by $ \rho^{A}_{i}=\frac
1{p_{i}}\mathrm{Tr}_{B}[(I^{A}\otimes
\Pi^{B}_{i})\rho^{AB}(I^{A}\otimes \Pi^{B}_{i})].$ Here $p_{i}$ is
the measurement probability for the $i-th$ local projector. Then the
classical correlation in the bipartite quantum state $\rho^{AB}$ can
be given by
\begin{equation}
C(\rho^{AB})=S(\rho^{A})-\sup_{\{ \Pi^{B}_{i}\}}S(\rho^{A|B}),
\end{equation}
where $S(\rho^{A|B})=\sum_{i=0,1}p_{i}S(\rho^{A}_{i})$ is the
conditional entropy of subsystem $A$ and $\sup(\cdot)$ signifies the
minimal value of the entropy with respect to a complete set of local
measurements. Quantum discord is simply defined by $
D(\rho^{AB})=I(\rho^{AB})-C(\rho^{AB}). $

According to Ref. \cite{Luo08}, the quantum mutual information for
Bell-diagonal mixed states $\rho^{AB}(t)$ in Eq. (4) is given by
\begin{equation}
I(\rho^{AB})=2+\sum_{i=1}^{4}\lambda_{i}(t)\log_{2}\lambda_{i}(t),
\end{equation}
and the classical correlation
\begin{equation}
C(\rho^{AB})=\sum_{j=1,2}\frac {1+(-1)^j
\Lambda_{max}(t)}{2}\log_{2}[1+(-1)^j \Lambda_{max}(t)].
\end{equation}
Here $\Lambda_{max}(t)=\max \{|c_x(t)|,|c_y(t)|,|c_z(t)| \}$. To
obtain the interesting behavior, we focus on three kinds of initial
states.

If the initial states $\rho^{AB}(0)$ satisfy the relation of
$c_x=\mp c_z$ and $c_y=\pm c_{x}^{2}$, the mutual information is
expressed by
\begin{equation}
I(\rho^{AB})=\sum_{j=1,2}[1+(-1)^j p(t)c_x]\log_{2}[1+(-1)^j
p(t)c_x],
\end{equation}
and the classical and quantum correlations are given by
\begin{equation}
D=C=\frac {I(\rho^{AB})}{2}
\end{equation}
This point means that the decoherence of quantum correlations is
identical to that of classical correlations. Because the classical
correlations are easily measured in the practical situations, the
behavior of quantum correlations can also be observed indirectly.
Figure 1 clearly shows the case of $D=C$. When the kernel of the
environments satisfies $A=a=\gamma$, the exponential decay of
classical and quantum correlations is plotted in Fig. 1(a). Compared
to the Markovian case, the decline of quantum and classical
correlations is slow and delayed to a large time interval. It is
also proven that the environment memory effects can yield some
positive results for the realization of quantum information
technology owing to the control of the system-reservoir
interactions. When the kernel is changed to be $A=10a\gg \gamma$,
the values of $D=C$ are decaying sinusoidally and the continuous
changes between nonzero values and zero ones are illustrated in Fig.
1(b).

If another one kind of initial states are chosen to be $c_x=\pm c_y$
and $c_z=\mp 1$, the quantum mutual information is obtained by
\begin{align}
I(\rho^{AB})=&\sum_{j=1,2}\frac {[1+(-1)^j
p(t)c_x]}{2}\log_{2}[1+(-1)^j p(t)c_x]&\\
\nonumber &+\sum_{j=1,2}\frac {[1+(-1)^j p(t)]}{2}\log_{2}[1+(-1)^j
p(t)]&.
\end{align}
The second term of the above equation always coincides with the
classical correlation $C(\rho^{AB})$. It is found that the change
rate of quantum correlation is proportional to that of classical
correlation decided by the decaying rate of the environment $p(t)$.
This synchronous phenomenon of quantum and classical correlations is
plotted in Figure 2. It is clearly seen that the values of classical
correlations are always larger than those of quantum correlations.
The decreasing and increasing of $C$ and $D$ simultaneously happen.

The initial states of $c_x=\pm c_z$ are used to reveal the sudden
changes of quantum and classical correlations. From Figure 3, it is
shown that there exists one certain characteristic time $t_c$ where
the decaying rates of both quantum and classical correlations vary
suddenly. To study the sudden changes, we analytically derive the
classical correlation. It is found that
\begin{widetext}
\begin{align}
C(\rho^{AB})=&\sum_{j=1,2}\frac {[1+(-1)^j
p(t)c_x]}{2}\log_{2}[1+(-1)^j p(t)c_x], &  |p(t)|< |p(t_c)|,& \\
\nonumber C(\rho^{AB})=&\sum_{j=1,2}\frac {[1+(-1)^j
p^2(t)c_y]}{2}\log_{2}[1+(-1)^j p^2(t)c_y], &  |p(t)|> |p(t_c)|,&
\end{align}
\end{widetext}
where $ |p(t_c)|=\frac {|c_x|}{|c_y|}$. According to the expressions
of $C(\rho^{AB})$, it is clearly shown that the decay of the
classical correlation suddenly changes at the characteristic time
$t_c$.

It is known that the sudden changes of quantum and classical
correlations are caused by the variation of the inherent properties
of the states. The distances from a given state to the closest state
are drawn on to understand the physical origin of the sudden
changes. According to Ref. \cite{Modi10}, the distance between the
state $\rho^{AB}$ and the closest classical state $\rho^{AB}_{cl}$
can quantify the quantum correlations defined by the relative
entropy
\begin{equation}
D(\rho^{AB}||\rho^{AB}_{cl})=\mathrm{Tr}(\rho^{AB}
\log_{2}\rho^{AB})-\mathrm{Tr}(\rho^{AB} \log_{2}\rho^{AB}_{cl}).
\end{equation}
For Bell-diagonal states,
$D(\rho^{AB}||\rho^{AB}_{cl})=D(\rho^{AB})$. It is the fact that the
sudden change of quantum correlation is closely dependent on the
variation of the closest classical state. For $t<t_c$, the closest
classical state  can be written by
\begin{align}
\rho^{AB}_{cl}=&(\lambda_1+\lambda_4)(|eg\rangle \langle
eg|+|ge\rangle \langle ge|)& \\ \nonumber
&+(\lambda_2+\lambda_3)(|ee\rangle \langle ee|+|gg\rangle \langle
gg|).
\end{align}
While the closest classical state for $t>t_c$,
\begin{align}
\rho^{AB}_{cl}=&\frac {\lambda_1+\lambda_2}{2}(|\Psi^{+}\rangle
\langle \Psi^{+}|+|\Phi^{+}\rangle \langle \Phi^{+}|)& \nonumber \\
&+\frac {\lambda_3+\lambda_4}{2}(|\Psi^{-}\rangle \langle
\Psi^{-}|+|\Phi^{-}\rangle \langle \Phi^{-}|).
\end{align}
Via the control of the kernel, the values of quantum correlations
are greater than those of classical correlations in the large time
interval from Fig. 3(a) and 3(b). When the kernel satisfies
$A=a=\gamma$, the characteristic time is obtained by $t_c=\ln \frac
{1+\sqrt{1-|c_x|/|c_y|}}{|c_x|/|c_y|}$ and numerically plotted in
Fig. 3(c). It is seen that the characteristic time can be enlarged
by the increasing of the ratio $|c_x|/|c_y|$. For the time interval
$t<t_c$, quantum correlation decline more slowly than classical
correlation.

\section{Discussion}

The decoherence of quantum and classical correlations is
investigated in the condition that two noninteracting qubits are
asymmetrically subjected to different non-Markovian Pauli channels.
It is found that the destruction of quantum and classical
correlations can be delayed via the engineering of the kernel
characterizing the environment memory effects. The dynamical
behavior of quantum and classical correlations are dependent on the
selection of the initial states. For one class of Bell-diagonal
mixed states, quantum correlations measured by quantum discord vary
synchronously with classical correlations. This phenomena provide us
a convenient way to evaluate quantum correlations by the direct
observation of classical correlations. Meanwhile, the decaying rates
of quantum correlations suddenly change at the characteristic time.
For $t<t_c$, quantum correlation decline more slowly than classical
correlation. In a large time interval, classical correlations are
less than quantum correlations which is useful for the realization
of quantum information processing. The physical explanation of
sudden changes of quantum correlations is obtained by the distance
between the quantum state and the closest classical state.

\section{Acknowledgements}

The work was supported by the Research Program of Natural Science
for Colleges and Universities in Jiangsu Province Grant No.
09KJB140009 and the National Natural Science Foundation Grant No.
10904104.

\newpage

{\Large Figure caption}

Figure 1

The dynamics of correlations are plotted when the initial state is
chosen to be $c_x=-c_z=c^{1/2}_{y}=0.6$. (a). The parameters of the
kernel $A=a=\gamma$; (b). the kernel satisfies $A=10a\gg \gamma$.
The solid lines denote quantum and classical correlations and the
dotted line represents the dynamics of correlations in the Markovian
case.

Figure 2

The dynamics of correlations are plotted when the initial state is
chosen to be $c_x=c_y=0.6$ and $c_z=-1$. (a). The parameters of the
kernel $A=a=\gamma$; (b). the kernel satisfies $A=10a\gg \gamma$.
The solid lines denote quantum correlations and the dashed lines
represent classical correlations.

Figure 3

(a). The dynamics of correlations are plotted when the initial state
is chosen to be $c_x=c_z=0.1$ and $c_y=0.16$ for the kernel
satisfying $A=a=\gamma$. (b). The dynamics of correlations are
plotted when the initial state is chosen to be $c_x=c_z=0.1$ and
$c_y=0.16$ for the kernel satisfying $A=10a\gg \gamma$. (c). The
characteristic time $t_c/a$ is plotted with the change of $c_y$ when
$c_x=c_z=0.1$ and $A=a=\gamma$.

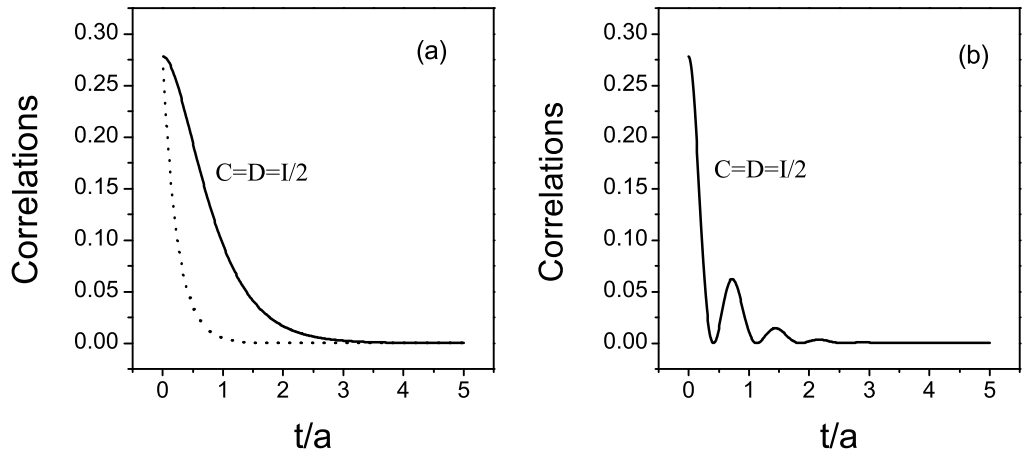
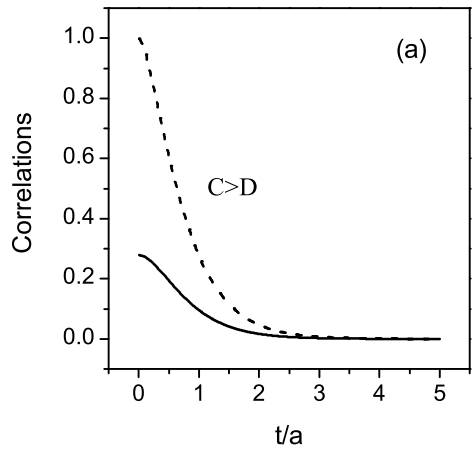
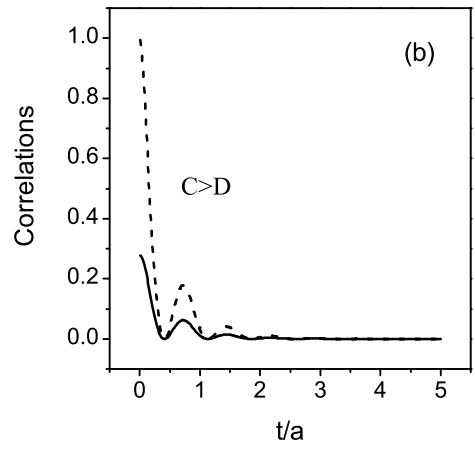
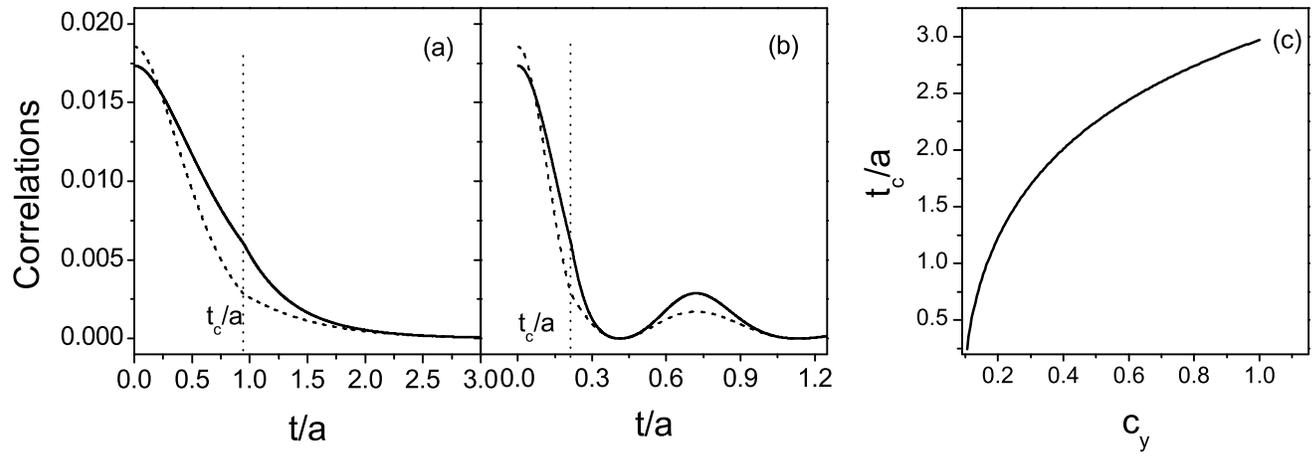
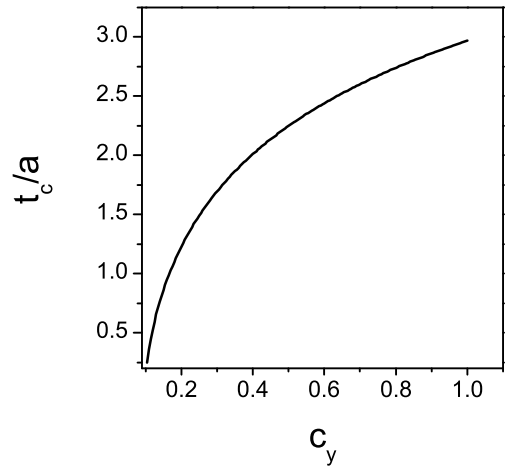
\end{document}